\documentclass{article}
\usepackage[sumlimits,intlimits,namelimits]{amsmath}
\usepackage{amssymb}
\usepackage[english]{babel}
\usepackage{bbm}
\usepackage{parskip}

\newcommand{\be}{\begin{eqnarray}}
\newcommand{\ee}{\end{eqnarray}}

\newcommand{\lb}{\label}
\def\>{\rangle}
\def\<{\langle}

\begin{document}
%
\begin{titlepage}
\hfill{LMU-ASC 40/08}
\vspace*{.5cm}
\begin{center}
{\large{{\bf Quasi-Normal Modes for Logarithmic Conformal Field Theory}}} \\
\vspace*{1.5cm}
Ivo Sachs\footnote{{\tt
Ivo.Sachs@physik.uni-muenchen.de}}\\
{\it Arnold Sommerfeld Center for Theoretical Physics (ASC),
Ludwig-Maximilians Universit\"{a}t\\
Theresienstrasse 37\\
D-80333, M\"{u}nchen\\
Germany}\\
\end{center}
\vspace*{1cm}
\begin{abstract}
\noindent {We generalize the one-to one correspondence between quasi-normal modes in 3- dimensional anti deSitter black holes and the poles of the retarded correlators in the boundary conformal field theory to include logarithmic operators in the latter. This analysis is carried out explicitly for the logarithmic  mode in topologically massive gravity at the chiral point. 
\\
\vspace*{.25cm}
}
\end{abstract}
\vspace*{.25cm}
\end{titlepage}
\section{Introduction}
It was observed in \cite{BSS1} that there is a one to one correspondence between the quasi-normal frequencies of linear perturbations in a 3-dimensional BTZ black hole background and the poles, in momentum space, of the retarded propagator of the respective dual operators in the boundary conformal field theory. This correspondence is conjectured to hold also in higher dimensions as a consequence of the AdS/CFT correspondence. This has lead to quantitative predictions about the hydrodynamic regime of strongly coupled large N supersymmetric quantum field theory \cite {Policastro:2001yc}. 

The observation made in \cite{BSS1} can be generalized to tensor perturbations when considering toplogically massive gravity \cite {Deser:1982vy} within the AdS/CFT correspondence, either as a theory of pure gravity \cite{Witten07} or embedded into string theory \cite{Kraus:2005zm,solo}.  For generic values of the mass $m$ the graviton becomes propagating in $AdS_3$ and the corresponding quasi-normal mode spectrum has been constructed recently in \cite {SS1}. For $m=1$, the chiral point, the modes found in \cite {SS1} are pure gauge and by themselves presumably have no physical interpretation.  However, it was pointed out in \cite{Carlip,Gru} that, at the chiral point, a new logarithmic solution arises which is propagating. 

In this note we construct the infinite tower of quasi-normal modes corresponding to this new mode and discuss its relation with the poles of the retarded correlators in logarithmic conformal field theory\footnote{The relevance of this mode was emphasized  to us by Daniel Grumiller.}. The key observation is that the presence of logarithmic solutions in the quasi-normal mode spectrum is related to the existence of double poles in the momentum representation of the retarded 
correlation function of the corresponding logarithmic operator in conformal field theory. This relation appears to be generic and can be understood in terms of a simple reconstruction formula of quasi-normal modes  presented below. Nevertheless, in view of the physical consistency of the specific model considered here we should mention that topologically massive gravity without truncation may lead to a pathological quantum theory since either the linearized graviton perturbation, or the BTZ black holes have been found to have negative energy depending on which sign one chooses for Newton's constant \cite{Strominger, Carlip, Gru}. However, these issues do not affect the correspondence established in this paper. 

\section{Quasi-Normal Modes }
\subsection{Algebraic structure}
There is a simple algebraic structure relating the massive graviton solution for generic mass $m>1$ to the logarithmic solution at $m=1$. To describe it we consider the equation for motion for tensor linear perturbations $h_{\mu\nu}$ in the transverse trace-less gauge 
 \be (\nabla^2+2)\left[\epsilon_\mu^{\
\alpha\beta}\nabla_\alpha h_{\beta \nu} +mh_{\mu\nu}\right] =0~~.
\lb{lin2} \ee
Introducing the operator $D_m$ through
\be
D_mh_{\mu\nu}=\epsilon_\mu^{\ \alpha\beta}\nabla_\alpha h_{\beta
\nu} +mh_{\mu\nu}\lb{Dm} \;,
\ee 
the 3rd order equation (\ref{lin2}) can be written as the product of three commuting first
order operators \cite{Strominger}
\be D_{+1}D_{-1}D_m h_{\mu\nu}=0\lb{3rd} \;.
\ee 
The solutions of $ D_{-1}D_{+1}h_{\mu\nu}=0$ describe a massless graviton in $2+1$ dimensions which is known not to propagate. Suppose now that $h_{\mu\nu}(m)$ is a solution to the first order equation 
\begin{equation}\label{first}
D_mh_{\mu\nu}=0~~.
\end{equation}
For $m=1$ this solution becomes degenerate with the massless graviton and thus becomes pure gauge. One way to see this is to note that  
\be 
D_{-m}D_m h_{\mu\nu}=(\nabla^2+3-m^2)h_{\mu\nu}=0~~. \lb{massive} 
\ee
One then shows \cite{SS1} that the change in the Riemann tensor produced by the metric perturbation
 satisfying (\ref{massive}) is given by
\be \delta R^{\mu\nu}_{\ \ \alpha\beta}={(1-m^2)\over
2}(h^\mu_\alpha \delta^\nu_\beta+h^\nu_\beta
\delta^\mu_\alpha-h^\mu_\beta
\delta^\nu_\alpha-h^\nu_\alpha\delta^\mu_\beta ) \;.\lb{change} \ee
On the other hand, it was shown in \cite{Gru} that for $m=1$ a new logarithmic solution $\tilde h_{\mu\nu}$ to (\ref{lin2}) appears with
\be
\tilde h_{\mu\nu}=\partial_m h_{\mu\nu}(m)|_{m=1}\;.
\ee
To continue we note that $\partial_m$ commutes with all geometric differential operators that do not depend on $m$. This implies, in particular, that 
\be 
\nabla^\mu\tilde h_{\mu\nu}=0\; .
\ee
Furthermore we can recover $h_{\mu\nu}$ in terms of $\tilde h_{\mu\nu}$ since
\be
D_1\tilde h_{\mu\nu}=-h(1)_{\mu\nu} \ .
\ee
The change in the Ricci tensor produced by $\tilde h_{\mu\nu}$ is then given by 
\be \delta R^{\mu}_{\;\;\;\nu}=-h^\mu_{\;\;\nu} \;.\lb{changet} \ee
which shows, in particular, that $\tilde h_{\mu\nu}$ is not pure gauge.

\subsection{Logarithmic quasi-normal modes}
It was shown in \cite{SS1} that in topologically massive gravity the quasi-normal modes for massive gravitons in the BTZ black hole background with metric ($u=\tau+\phi$, $v=\tau-\phi$)
\begin{equation}\label{guv}
ds^2 = \frac{1}{4}\left(du^2-2\cosh(2\rho)dudv+dv^2\right)+d\rho^2\;,
\end{equation}
are descendents of a "chiral highest weight" solution, $h(m)_{\mu\nu}$,  to the first order equation (\ref{first}) satisfying $L_1h_{\mu\nu}=0$ or $\bar L_1h_{\mu\nu}=0$ (but not both). 
The generators $L_{\pm 1}$ and $L_0$ form  a representation of the Lie algebra $SL(2,R)$,
\be\label{alg} 
[L_0, L_{\pm 1} ]=\mp L_{\pm
1}~,~~[L_1,L_{-1}]=2L_0~~.  
\ee 
in the black hole background through
\begin{eqnarray}
L_0&=&-\partial_u\nonumber\\
L_{-1}&=&e^{-u}\left(-\frac{\cosh(2\rho)}{\sinh(2\rho)}\partial_u-\frac{1}{\sinh(2\rho)}
\partial_v-\frac{1}{2}\partial_\rho\right)\lb{Kf}\\
L_{1}&=&e^{u}\left(-\frac{\cosh(2\rho)}{\sinh(2\rho)}\partial_u-\frac{1}{\sinh(2\rho)}\partial_v+
\frac{1}{2}\partial_\rho\right)\nonumber\;.
\end{eqnarray}
Similarly, $\bar L_0,\bar L_{1},\bar L_{-1}$ are obtained from 
(\ref{Kf}) by substituting $u\rightarrow v$ and $v\rightarrow u$.

For $L_1h_{\mu\nu}=0$ the descendents
$$
h(m)^{(n)}_{\mu\nu}=(L_{-1}\bar L_{-1})^nh(m)_{\mu\nu}\;.
$$
generate the complete tower of quasi-normal modes for fixed $m$  with left-moving quasi-normal frequencies
 \be
\omega^L_n=-k-2i(h_L(m)+n)~,~~n\in N \;, 
\ee
and $ h_L(m)=\frac{m}{2}-\frac{1}{2}$. Similarly, $\bar L_1h_{\mu\nu}=0$ leads to the quasi-normal mode spectrum for the right-moving quasi-normal frequencies. For $m=1$, however, the solution of (\ref{first}) becomes pure gauge an can thus not be a quasi-normal mode. We will now show that at the chiral point the descendents of $\tilde h_{\mu\nu}$ form a infinite tower of quasi-normal modes instead.  

It is clear from the previous subsection that $L_1\tilde h_{\mu\nu}=0$. Thus $\tilde h_{\mu\nu}$ is chiral highest weight. Furthermore, since $L_k$ and $\bar L_k$ commute with the equation of motion \cite{SS1}, 
$$
\tilde h^{(n)}_{\mu\nu}=(\bar L_1 L_{-1})^{(n)}\tilde h_{\mu\nu}
$$ 
satisfies the third order equation (\ref{lin2}). 
All modes are ingoing at the horizon since $h_{\mu\nu}$ has this property. Finally all modes, except the highest weight mode $\tilde h_{\mu\nu}$ itself, fall-off exponentially in time and large radial distances. Concretely we have\footnote{In what follows we assume vanishing angular momentum for the quasi-normal modes to avoid clutter. Including it is not a principle obstacle.} \cite{Gru}
\be
\tilde h_{\mu\nu}=-y(\tau,\rho)\psi_{\mu\nu}\;,\lb{n0}
\ee
where $y(\tau,\rho)=\tau+\log[\sinh(\rho)]$, and \cite{SS1}
\begin{equation}
\psi_{\mu\nu}=\begin{pmatrix} 0&0&0\cr
        0&1&\frac{2}{\sinh(2\rho)}\cr
        0&\frac{2}{\sinh(2\rho)}&\frac{4}{\sinh^2(2\rho)}
        \end{pmatrix}~~.
\end{equation}
Since this mode is growing in time and $\rho$ one might be tempted to disqualify it as a quasi-normal mode. On the other hand, all components of the corresponding perturbation of the Ricci-tensor  
\be \delta R^{\mu}_{\;\nu}=-\psi^\mu_{\;\;\nu}\;,\lb{changeR} \ee
fall-off exponentially. It can be shown, furthermore that the energy flow, $\sqrt{|g|}T^\rho_{\;\;\tau}$ for this mode falls off exponentially in $\rho$. We postpone the discussion of the linear growth in $\tau$ to next section when discussing the relation with conformal field theory. 

The first descendent $\tilde h^{(1)}_{\mu\nu}$ takes the form
\be
\tilde h^{(1)}_{\mu\nu}=\left(\frac{1}{2}-y(\tau,\rho)\right)\psi^{(1)}_{\mu\nu}\;,\lb{n1}
\ee
where
\begin{equation}
\psi^{(1)}_{\mu\nu}=\bar L_{-1}L_{-1}\psi_{\mu\nu}=\frac{2 e^{-2\tau}}{\sinh^2(\rho)}\begin{pmatrix} 0&1&\frac{2}{\sinh(2\rho)}\cr
        1&1&\frac{2\cosh(\rho)}{\sinh(\rho)}\cr
        \frac{2}{\sinh(2\rho)}&\frac{2\cosh(\rho)}{\sinh(\rho)}&4\frac{1+2\cosh(2\rho)}{\sinh^2(2\rho)}        \end{pmatrix}\;.
\end{equation}
This is a genuine gravitational quasi-normal mode with exponential fall-off in $\rho$ and $\tau$. In view of a conformal field theory interpretation of $\tilde h^{(1)}$ we should note that the $vv$-component of the metric is not dominant at large $\rho$ which in turn leads to difficulties in identifying the dual operator in the CFT. The curvature perturbation induced by $\tilde h^{(1)}$ is then obtained using  (\ref{changet}) as
\be \delta R^{\mu}_{\;\nu}=-(\psi^{(1)})^\mu_{\;\;\nu}
 \;.\lb{changeR2} \ee
The structure of the higher modes $\tilde h^{(n)}_{\mu\nu}$ with $n>2$ is similar with $ e^{-2\tau}$ replaced by $ e^{-2n\tau}$. To summarize, topologically massive gravity at the chiral point has an infinite tower of quasi-normal modes with quasi-normal frequencies (restoring the $k$ dependence) 
 \be
\omega^L_n=-k -2in~,~~n\in N \lb{l} \;.
\ee
A qualitatively new feature of these modes is the appearance of a linear dependence in time of (\ref{n0},\ref{n1}) on top of the 
exponential decay in time. In the next section we will see that this feature has a natural explanation in logarithmic conformal field theory. The same analysis can be done for the
right-moving modes found in \cite{SS1} producing the second set of
the quasi-normal modes.

\section{Relation to Logarithmic CFT}
The simplest version of a logarithmic conformal field theory (see e.g. \cite{Flohr:2001zs} for a review), which is sufficient for our purpose arises
in the presence of two operators $C$ and $D$ with degenerate eigenvalue of $L_0$ such that
\be
L_0 |C>=h|C>\;,\qquad L_0|D>=h|D>+|C>\lb{c1}\;.\lb{lcdef}
\ee 
The 2-point functions of these operators are then given by 
\begin{eqnarray}
<C(x) C(0)>&=&0\nonumber\\
<C(x) D(0)>&=&\frac{c}{x^{2h}}\lb{lcor}\\
<D(x) D(0)>&=&\frac{1}{x^{2h}}\left[a-2c\log(x)\right]\nonumber\;,
\end{eqnarray}
respectively. Note that (\ref{c1}) does not uniquely fix $C$ and $D$. In particular $D'=D+\alpha C$ satisfies (\ref{c1}). 
This freedom can be used to adjust the constant $a$ to any suitable value. 

In view of the quasi-normal modes we will be interested in the location an the nature of the poles, in momentum space, of the retarded correlators $G^{CC}_R(t,\sigma)$, $G^{CD}_R(t,\sigma)$ and $G^{DD}_R(t,\sigma)$ in finite temperature conformal field theory.  Now, $G^{CD}_R(t,\sigma)$ is identical with that of  the two point function in ordinary conformal field theory. Its momentum space representation can thus be inferred from that of the commutator whose pole structure is that of (see \cite{BSS1}  for details)
\begin{eqnarray} 
&&\bar{\cal D}^{DC}(p_+) \propto 
\Gamma\left(h_L+ip_+\right) \Gamma\left(h_L-ip_+\right) \ , 
\label{9} 
\end{eqnarray} 
where $p_\pm\!=\!\frac{1}{2}(\omega\pm k)$. This function has poles  
in both the upper and lower half of the $\omega$-plane.  
The poles lying in the lower half-plane are the same as the poles of the  
retarded correlation function $G^{CD}_R(t,\sigma)$. 
Restricting the poles of (\ref{9}) to the lower half-plane, 
we find two sets of simple poles 
\begin{eqnarray} 
\omega_L&=&-k-2 i (n+h_L) \;. 
\label{10} 
\end{eqnarray} 
where $n$ takes the integer values $(n=0,1,2,...)$. 
This  set of poles characterises the decay 
of the perturbation on the CFT side. 

Next $G^{DD}_R(t,\sigma)$ can be inferred as in \cite{Lewis} noting that
\be
<D(x) D(0)>=\partial_h<C(x) D(0)>\;,
\ee
thus
\begin{eqnarray} 
\bar{\cal D}^{DD}(p_+)& \propto& 
\Gamma'\left(h_L+ip_+\right) \Gamma\left(h_L-ip_+\right) \nonumber\\ 
&&\qquad+\Gamma\left(h_L+ip_+\right) \Gamma'\left(h_L-ip_+\right)
 \;.
\end{eqnarray} 
Again, only the poles in the lower half plane are relevant. We then conclude that the momentum space representation of  $G^{DD}_R(t,\sigma)$ has double poles while that of $G^{CD}_R(t,\sigma)$ has simple poles at the same location. 

We will argue below that it is precisely these double poles that are responsible for the linear time dependence of the corresponding quasi-normal mode. Before doing so, however, we need to assign the bulk perturbation to the operators $C$ and $D$. This proceeds in close analogy with the analysis presented in \cite{Lewis}. For $m>1$ the tensor perturbation $h(m)_{\mu\nu}$ is dual to a non-degenerate boundary operator $C$ with conformal weight $h_L=\frac{m-1}{2}$. At the chiral point, $m=1$ 
the  perturbation $h(m)_{\mu\nu}$ becomes pure gauge. On the other hand (\ref{lcdef}) together with $L_1D=0$ imply for the corresponding bulk perturbation $\Phi_D$ \cite{Lewis} 
\be
\Phi_D=(y(\tau,\rho)+\alpha)h_{\mu\nu}
\ee
which is just the definition of $\tilde h_{\mu\nu}$. We then conclude that $\tilde h_{\mu\nu}$ is the bulk perturbation for the logarithmic partner of $D$ agreement with the conclusion reached in \cite{Gru}. 

Let us now finally explain the linear time dependence in $\tilde h_{\mu\nu}$. For this we notice that the lowest lying quasi-normal mode  can be reconstructed for a given boundary correlation function as
\be
h_{\mu\nu}=\psi_{\mu\nu}\oint\limits_{C} \;d\omega e^{-i(\omega \tau+ k\phi)} f(\rho) \bar{\cal D}(p_+)\lb{rec}
\ee
where the contour is around the pole of $\bar{\cal D}$ closest to the real axis and the function
$f(\rho)=\sinh(\rho)^{-i\omega}\cosh(\rho)^{-ik}$ determines the extension of $h_{\mu\nu}$ into the bulk\footnote{This formula  applies equally to the reconstruction of the lowest lying scalar quasi-normal mode in \cite{SS1} by replacing $\psi_{\mu\nu}$ by the identity.}. Now, if $\bar{\cal D}(p_+)$ has a simple pole then (\ref{rec}) reproduces the tensor perturbation $h_{\mu\nu}$. On the other hand, if  $\bar{\cal D}(p_+)$ has a double pole as is the case for the logarithmic operator $D$, then (\ref{rec}) reproduces $\tilde h_{\mu\nu}$ with just the right dependence on $\tau$ and $\rho$. This shows that the linear time dependence in $\tilde h_{\mu\nu}$ is related to the fact that the spectral density of the retarded boundary correlation function has a double pole. In particular, the highest weight mode (\ref{n0})  which grows linearly in time can be thought of a "quasi-normal mode" with zero quasi-normal frequency corresponding to a double pole on the real line.


\subsection*{Acknowledgments:}
This was work supported in parts by the Transregio TRR~33 `The Dark
Universe',  the Excellence Cluster `Origin and Structure of the
Universe' of the DFG as well as the DFG grant  Ma 2322/3-1. I would like to thank D. Grumiller for helpful correspondence and Sergey Solodukhin for helpful comments and for pointing out some mistakes in an earlier version of the manuscript.


\end{document}